\documentclass{article}
\usepackage{amsmath}
\usepackage{graphicx}
\usepackage{enumerate}
\usepackage{natbib}
\usepackage{url} 
\usepackage{fullpage}
\usepackage[utf8]{inputenc}
\usepackage{amsmath,amsfonts,amssymb,amsthm}
\usepackage{tikz}
\usetikzlibrary{calc,patterns,angles,quotes}

\newcommand{\vc}[1]{\mathbf}
\newcommand{\mat}[1]{\mathbf}

\DeclareMathOperator{\var}{Var}

\DeclareMathOperator{\mse}{MSE}

\DeclareMathOperator{\tr}{tr}

\def\T{{ \mathrm{\scriptscriptstyle T} }}

\newtheorem{theorem}{Theorem}
\newtheorem{lemma}{Lemma}

\newtheorem{definition}{Definition}

\theoremstyle{remark}
\newtheorem{remark}{Remark}

\newtheorem{corollary}{Corollary}

\title{\bf Penalized angular regression \\ for personalized predictions}
  \author{Kristoffer H. Hellton\\
    Department of Mathematics, University of Oslo, Norway \\
   Norwegian Computing Center, Norway.}

\begin{document}

\maketitle

\begin{abstract}
Personalization is becoming an important feature in many predictive applications. We introduce a penalized regression method implementing personalization inherently in the penalty. Personalized angle (PAN) regression constructs regression coefficients that are specific to the covariate vector for which one is producing a prediction, thus personalizing the regression model itself. This is achieved by penalizing the angles in a hyperspherical parametrization of the regression coefficients. For an orthogonal design matrix, it is shown that the PAN estimate is the solution to a low-dimensional eigenvector equation. Using a parametric bootstrap procedure to select the tuning parameter, simulations show that PAN regression can outperform ordinary least squares and ridge regression in terms of prediction error. We further prove that by combining the PAN penalty with an $L_{2}$ penalty the resulting method will have uniformly smaller mean squared prediction error than ridge regression, asymptotically. Finally, we demonstrate the method in a medical application. 
\end{abstract}
\noindent%

\section{Introduction} 
The ambition to perform personalization when predicting is becoming an important feature of many applications: medicine \citep{cheng2012molecular,carrion2016personalized}, marketing \citep{tang2013prediction}, item recommendation \citep{rafailidis2014content}, nutrition \citep{zeevi2015personalized}, education \citep{reber2018personalized} and fraud detection \citep{cama2018fraud}; all applications targeting the individual. Personalized medicine or precision medicine, for instance, utilizes the genomic information, proteins, or the environment of a patient to predict individualized treatment decisions 
\citep{hamburg2010path,zhang2017personalized}. Other examples include personalized marketing, delivering individualized product prices or messages to specific costumers, and item recommendation, predicting the rating of an item or product for a given user. These applications call for statistical prediction methods targeting the individual also on the \emph{methodological level}, meaning that the estimated model itself may vary with each prediction one wishes to make. The aim is to minimize the prediction error for each individual covariate vector, instead of minimizing the average prediction error. We propose a form of penalized regression which inherently features this personalized approach to prediction.

Penalized regression is a class of methods particularly useful for prediction in high-dimensional or multicollinear data. The typical methods, e.g.\ ridge regression, lasso and elastic net \citep{hoerl1970ridge,tibshirani1996regression,zou2005regularization} penalize some norm of the regression coefficients such as the $L_{1}$, $L_{2}$ or $L_{p}$ norm, or some combination thereof. The norms typically have a geometric interpretation; the $L_{2}$ norm, for instance, equals the Euclidean length of the regression coefficient vector. Hyperspherical coordinates parametrize a $p$-dimensional vector geometrically in terms of its length and $p-1$ angles and generalize polar coordinates to $p$ dimensions. Hyperspherical coordinates are commonly used in physics, e.g.\ to solve three- and four-particle problems and the Laplace's equation \citep{ohrn1983hyperspherical,cohl2011laplace}. There has been an increased interest in the statistical distribution of angles in high dimension \citep{cho2009inner,cai2013distributions} and the use of the hyperspherical parametrization in statistics and machine learning \citep{pourahmadi2015distribution,liu2017deep}. Related fields also include directional statistics, regression models for circular and spherical outcomes \citep{mardia1972statistics}, and compositional data \citep{scealy2011regression}. 

In the context of model selection, \citet{claeskens2003focused} introduced the concept of addressing the \emph{aim} of the statistical analysis with the focused information criterion (FIC). The focused model selection approach defines an \emph{a priori} quantity-of-interest or focus parameter to guide the selection of a statistical model, instead of considering overall goodness-of-fit measures \citep{claeskens2008model}. For different aims, or foci, different models may then be selected. Other frameworks also introduce notions of a pre-defined target parameter representing the scientific question, such as targeted learning \citep{van2011targeted}. The focused approach was extended to include a specific prediction as the aim by \citet{hellton2018fridge}, framing the resulting model as personalized.

Currently, the term personalization is typically understood as standard regression models, where covariates account for the differences and heterogeneity between individuals \citep{tian2015statistical,kosorok2019precision}. Personalization of the regression model, however, can be implemented in several ways in the penalized regression context. \citet{hellton2018fridge} and \citet{huang2019tuning} proposed to vary the tuning parameter in ridge regression with each covariate vector, $x_0$, for which one wishes to make a prediction. The personalized tuning parameter, $\lambda_{x_0}$, can be estimated by minimizing the prediction error of $x_0$ via a two-stage plug-in procedure or adaptive validation. In this paper, we instead incorporate the personalization in the penalty structure itself. To personalize a prediction, adapting it to a specific individual, was shown to be inherently connected to the angle between the regression coefficients and the covariate vector in question. We therefore use the focused approach combined with a hyperspherical parametrization to achieve a guided penalization of the regression coefficients. The resulting method produces personalized regression coefficients and predictions, implementing personalization at the methodological level. 

In a personalized framework, one aims to make inference regarding a single, specific case which has been and may only be observed once. The advantage of personalization therefore relies on the structure and, in particular, the heterogeneity of the data. \citet{liu2016there} commented: ``The costs of individualization often outweighed its benefits'', but that highly heterogeneous data will benefit more from personalization than homogeneous data. This highlights the opportunity of the Big Data era where data are becoming more heterogeneous. Big Data are typically characterized by a large sample size aggregated from multiple data sources and at different times, creating an intrinsic heterogeneity \citep{fan2014challenges}. This heterogeneity can be exploited by personalized prediction methods. 

The remainder of the paper is organized as follows: In Section 2 we present the personalized angle penalty and show that it penalizes the angle parameter in a hyperspherical parametrization of linear regression. In Section 3, the new penalty is combined with ridge regression. Section 4 presents a simulation study comparing personalized angle regression to ridge and OLS regression and in Section 5, we illustrate the method in a medical application. Concluding remarks are discussed in Section 6, and all proofs are collected in the Appendix. 


\section{Personalized angle regression} 
\subsection{Definition}
Suppose we have observed data $\{y_{i},x_{i}\}, i = 1, \dots, n$, consisting of $p$-dimensional covariate vectors, $x_{i} \in \mathbb{R}^p$, and univariate continuous outcomes, $y_{i}\in\mathbb{R}$, and consider the linear regression model
$$y_{i}= x_{i}^{\T}\beta+ \varepsilon_{i} \quad i=1,\dots,n,$$ 
where $\beta\in \mathbb{R}^p$ is a $p$-dimensional vector of regression coefficients and $\varepsilon_i \in \mathbb{R}^n$ is an identically and independently distributed noise term with zero mean, $E(\varepsilon_{i}) =0$, and variance, $\var(\varepsilon_{i}) =\sigma^{2}$. 
The vector of outcomes is denoted $Y = [y_1, \dots, y_n]^{\T}$, and $X$ denotes the $n \times p$ design matrix with $x_i^{\T}$ as each row. The design matrix and outcome vector are assumed to be centered. 


In a personalized prediction context, the primary aim is optimal predictive ability. We propose to penalize the prediction for a given covariate vector, $x_{0}$, to improve the prediction error specifically, ignoring the estimation error. This will leverage the heterogeneity in the covariates to personalize the regression model. The covariate vector $x_0$ may or may not be in $X$ and represents an instance for which we wish to produce a prediction, e.g.\ a patient in the personalized medicine context. The personalization of the regression model requires the regression coefficients to be recalculated for each new prediction. The penalty we introduce is based on the normalized inner product between $x_0$ and $\beta$ and has an intuitive interpretation in hyperspherical coordinates. Therefore, as the resulting regression estimates are optimal for the specific $x_{0}$, we term the method Personalized Angle (PAN) regression. 

\begin{definition}[Cartesian coordinates]\label{def:cartesian}
The Personalized Angle (PAN) estimator for a specific covariate vector $x_{0}$, $\hat \beta_{x_{0}} = (\hat \beta_{1}, \dotsm \hat \beta_{p})^{\T}$ is defined as 
\begin{equation}
\mathbf{\hat{\beta}}_{x_{0}}(\lambda) = \arg\min_{\beta}\left\{\sum_{i = 1}^n \left(  y_{i} - x_{i}^{\T}\beta \right)^2 + \frac{\lambda}{x_{0}^{\T}x_{0}}\;\frac{\beta^{\T}x_{0}x_{0}^{\T}\beta}{\beta^{\T}\beta}\right\}, \label{penRSS}
\end{equation}
where $\lambda \in \mathbb{R}$ is a tuning parameter.  
\end{definition}
\noindent The PAN regression penalizes the (squared) $L_2$ norm of the normalized prediction given $x_0$ 
$$ J(\beta) = \frac{\beta^{\T}x_{0}x_{0}^{\T}\beta}{x_{0}^{\T}x_{0}\beta^{\T}\beta} = \|\gamma_0^{\T}\gamma_{\beta}\|_2^2, $$
where $\|\cdot\|_2$ denotes the Euclidean $L_{2}$ norm and $\gamma_{\beta} = \beta/\|\beta\|_2$ and $\gamma_0 = x_{0}/\|x_{0}\|_2$ are the vectors $x_0$ and $\beta$ scaled to unit length. The penalty shrinks the prediction for $x_0$ towards zero which introduces a bias, but lowers the variance, with an optimal trade-off improving the mean squared prediction error of $x_{0}$ only. 

\begin{remark}
In the parameter space, the zero prediction corresponds to a hyperplane with $x_{0}$ as its normal vector, denoted by $H_{0}$: $$H_{0} = \{ \beta \in \mathbb{R}^{p}: x_{0}^{\T}\beta = 0 \},$$ 
with dimension, $ \dim(H_{0}) = p-1$. The penalty in Equation \eqref{penRSS} therefore shrinks the estimated regression coefficient vector towards the hyperplane $H_{0}$. As the tuning parameter, $\lambda$, increases, the part of the estimate orthogonal to $H_{0}$ decreases. In the limit $\lambda \to \infty$, the prediction becomes zero and the estimate converges to the projection of the OLS estimate, $\tilde \beta$, unto the hyperplane $H_{0}$. As the penalty function in Equation \eqref{penRSS} is bounded, the tuning parameter value may, however, also be \emph{negative}. This corresponds to shifting the prediction away from zero, and essentially ``expanding'' rather than shrinking the prediction. When later combined with ridge regression, this feature is key. 
\end{remark}

\subsection{Angular interpretation} 
Geometrically, any point $x_{i}$ can be described by a length $r$ and $p-1$ angles, $\alpha_{1},\dots, \alpha_{p-1}$, defined relative to the unit vectors. The standard hyperspherical parametrization, generalizing polar coordinates to $\mathbb{R}^p$, is given by
\begin{align*}
x_{i,1} &= r \cos \alpha_{1}, \\
x_{i,2} &= r \sin \alpha_{1}\cos \alpha_{2}, \\
& \vdots \\
x_{i,p-1} &= r \sin \alpha_{1} \sin \alpha_{2} \cdots \sin \alpha_{p-2} \cos \alpha_{p-1}, \\
x_{i,p} &= r \sin \alpha_{1} \sin \alpha_{2}   \cdots \sin \alpha_{p-2} \sin \alpha_{p-1},
\end{align*}
where $r\geq 0$ and the angles fulfill $0 \leq \alpha_i \leq \pi$ for $i =1,2,\dots,p-2$ and $-\pi < \alpha_{p-1} \leq \pi$. Using hyperspherical coordinates, we can reparametrize the regression coefficient vector 
\begin{equation}
\beta =  r_{\beta} \; \gamma_{\beta}, 
\label{eq:hyperspherical}
\end{equation}
by its length, $r_{\beta} = \|\beta\|_2$ and a direction vector, the normalized $\beta$ vector 
\begin{equation}
\gamma_{\beta}  = \left(\cos(\alpha_{\beta,1}), \dots,\; \sin (\alpha_{\beta,1}) \cdots \sin (\alpha_{\beta,p-2}) \sin (\alpha_{\beta,p-1})\right)^{\T}.
\label{eq:direction}
\end{equation}
In two dimensions, $p = 2$, we can transform standard linear regression into a nonlinear regression problem
\begin{align*}
y_{i} &= x_{i}^{\T}\beta+ \varepsilon_{i}  =  r_{\beta} r_{i} \left(\cos(\alpha_{i})\cos(\alpha_{\beta}) + \sin(\alpha_{i})\sin(\alpha_{\beta}) \right)  + \varepsilon_i, \\
&=r_{\beta} r_{i} \cos (\alpha_{i}- \alpha_{\beta})  + \varepsilon_i, \qquad i=1,\dots,n,
\end{align*}
where $r_{i}$ and $\alpha_{i}$ are the length and the angle of the $i$th covariate vector, respectively. The regression parameters can then be found by estimating the amplitude, $r_{\beta}$, and the phase shift, $\alpha_{\beta}$. This  reparametrization supplies an alternative estimation approach for the linear regression problem. In general dimension, the transformed model is estimated by minimizing the following residual sum-of-squares
 \begin{align} (\hat r_\beta, \hat \alpha_{\beta,1}, \dots,\hat \alpha_{\beta,p-1}) = \arg\min_{r_\beta, \alpha_\beta} \Bigg\{&\sum_{i = 1}^n \Big[  y_{i} - r_{\beta}r_{i} \big( \cos(\alpha_{\beta,p-1} -\alpha_{i,p-1})\prod_{j=1}^{p-2}\sin \alpha_{\beta,j} \sin \alpha_{i,j} \nonumber\\ 
& \qquad \quad + \sum_{j=2}^{p-2}\cos \alpha_{\beta,j}\cos\alpha_{i,j} \prod_{k=1}^{p-2}\sin \alpha_{\beta,k} \sin \alpha_{i,k} \big) \Big]^2
\Bigg\}, \label{eq:sphereRSS}  
 \end{align}
 which yields the ordinary least squares (OLS) estimate $\tilde\beta = (X^{\T}X)^{-1}X^{\T}Y$ transformed to hyperspherical coordinates 
\begin{align*}
 \tilde \alpha_{\beta,p-i}&=\arccos {\frac {\tilde\beta_{p-2}}{\sqrt {{\tilde\beta_{p}}^{2}+{\tilde\beta_{p-1}}^{2} + \cdots +{\tilde\beta_{p-i}}^{2}}}},\quad i = 1,\dots, p-2, \\
\tilde  \alpha_{\beta,p-1}&={\begin{cases}\arccos {\frac {\tilde\beta_{p-1}}{\sqrt {{\tilde\beta_{p}}^{2}+{\tilde\beta_{p-1}}^{2}}}}&\tilde\beta_{p}\geq 0\\[6pt]2\pi -\arccos {\frac {\tilde\beta_{p-1}}{\sqrt {{\tilde\beta_{p}}^{2}+{\tilde\beta_{p-1}}^{2}}}}&\tilde\beta_{p}<0\end{cases}}\,, \quad \tilde r_{\beta}={\sqrt {{\tilde\beta_{p}}^{2}+{\tilde\beta_{p-1}}^{2}+\cdots +{\tilde\beta_{2}}^{2}+{\tilde\beta_{1}}^{2}}}.
\end{align*}

\subsubsection{Penalizing the length}
Ridge regression \citep{hoerl1970ridge} adds a squared $L_{2}$ penalty to the residual sum-of-squares in Equation \eqref{eq:sphereRSS}, which corresponds to the squared length of the regression coefficient vector in hyperspherical coordinates:  
$$ J(\beta) = \| \beta\|^2_2 = \sum_{j=1}^{ p} \beta_{j}^{ 2} = r_{\beta}^{ 2}.$$
Ridge regression thus shrinks the length towards the origin. The ridge estimate has the explicit solution $\tilde\beta(\lambda) = (X^{\T}X +\lambda I_{p})^{ -1}X^{\T}Y$, where $I_p$ is the $p$-dimensional identity matrix and the tuning parameter $\lambda$ controls the penalization. Zero penalization corresponds to the OLS estimate, $\tilde \beta(0) = \tilde\beta$.

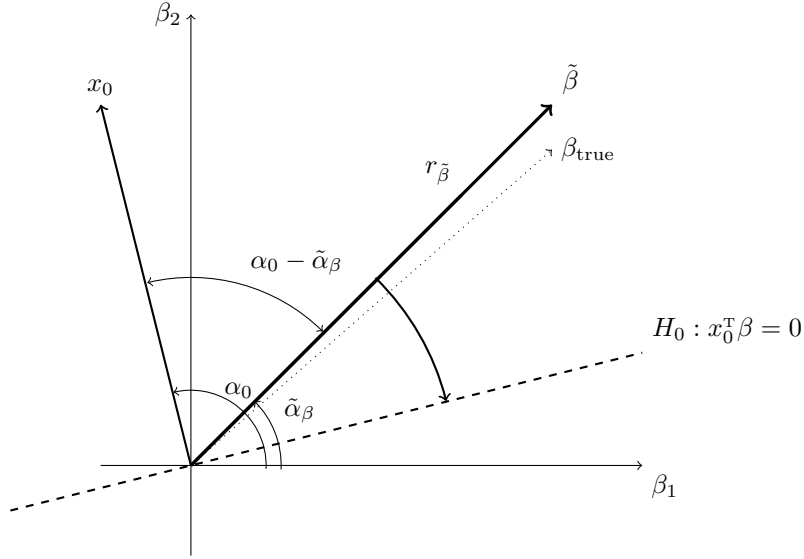
\begin{figure}%
\centering
\begin{tikzpicture}[scale=1.2]
\coordinate (origo) at (0,0);
\draw[->] (-1,0) -- (5,0) node[below right] (xaxis) {$\beta_{1}$};
\draw[->] (0,-1) -- (0,5) node[left] {$\beta_{2}$};
\draw[black,very thick,->] (0,0) -- (4,4) node (reg) [above right] {$\tilde\beta$};
\draw[black,very thick,-] (0,0) -- (3,3) node [above left] {$r_{\tilde\beta}$};
\draw[dotted,->] (0,0) -- (4,3.5) node[right] {$\beta_{\text{true}}$};
\draw[black,thick,->] (0,0) -- (-1,4) node (pred) [above] {$x_{0}$};
\draw[black,thick,dashed] (-2,-1/2) -- (5,5/4) node [above right] (orth) {$H_{0} : x_{0}^{\T}\beta = 0$};
\pic ["$\alpha_{0} - \tilde\alpha_{\beta}$",draw, <->, angle eccentricity=1, angle radius=2.5cm,above right] {angle = reg--origo--pred};
\pic ["$\alpha_{0}$",draw, ->, angle eccentricity=1, angle radius=1.0cm,above] {angle = xaxis--origo--pred};
\pic ["$\tilde\alpha_{\beta}$",draw, ->, angle eccentricity=1, angle radius=1.2cm,above right] {angle = xaxis--origo--reg};
\pic [draw, <-, angle eccentricity=1,angle radius=3.5cm,thick,black] {angle = orth--origo--reg};
\end{tikzpicture}
\caption{
Illustration of the shrinkage of the PAN penalty in two dimensions.}%
\label{angularOrigin}%
\end{figure}

\subsubsection{Penalizing the angle}
In two dimensions, $p=2$, the PAN penalty reduces to a squared cosine penalty on the angle parameters $\alpha_{\beta}$ and $\alpha_{0}$ 
\begin{align}
J(\beta) &=  \frac{\beta^{\T}x_{0}x_{0}^{\T}\beta }{x_{0}^{\T}x_{0}\beta^{\T}\beta} = \cos^{2}\left(\alpha_{\beta}-\alpha_{0}\right) =  1 - \cos^{ 2}\left(\alpha_{\beta}-\left(\alpha_{0} \pm \frac{\pi}{2}\right)\right), \label{eq:PANpenalty}
\end{align}
and hence corresponds to a ridge-type penalty on the \emph{angle parameter}. The shrinkage enforced by the PAN penalty therefore acts as a rotation of the OLS estimate. 

Figure \ref{angularOrigin} visualizes the OLS estimate, $\tilde\beta$, parametrized by its length $\tilde r_{\beta}$ and the angle $\tilde\alpha_{\beta}$ in the parameter space. The covariate vector, $x_{0}$, given by the angle $\alpha_{0}$, is visualized by overlying the covariate space over the parameter space. The zero prediction for $x_{0}$ is then used as an angular origin to shrink towards. The prediction equals zero when the regression coefficients, $\beta$, fulfills the equation $x_{0}^{\T}\beta = 0$, i.e.\ the vectors $\beta$ and $x_{0}$ are orthogonal. In two dimensions, this corresponds to the angle of $\beta$ being equal to $\alpha_{\beta} = \alpha_{0} \pm \frac{\pi}{2}$, visualized by the dashed line in Figure \ref{angularOrigin}. Hence, when $\lambda$ increases, the estimated angle rotates away from $\tilde\alpha_{\beta}$ as illustrated in Figure \ref{angularOrigin} towards $H_0$, the line orthogonal to $x_{0}$. The estimated angle, $\hat\alpha_{\beta}$, is rotated towards the closes of the two angles $\alpha_{0} \pm \frac{\pi}{2}$, shrinking the prediction towards zero. For a negative tuning parameter value, on the other hand, the estimated angle is rotated away from $H_0$ and towards $x_{0}$.  


With the PAN penalty in Equation \eqref{eq:PANpenalty}, the penalized residual sum-of-squares regularizing the angle parameter is given
$$ 
 (\hat r_{\beta, x_0}(\lambda),\hat \alpha_{\beta,x_0}(\lambda)) =\arg\min_{
r_{\beta},\alpha_{\beta}
  }\Bigg\{\sum_{i = 1}^n \big[  y_{i} - r_{\beta}r_{i}\left(\cos(\alpha_{i})\cos(\alpha_{\beta}) -  \sin(\alpha_{i})\sin(\alpha_{\beta})\right) \big]^2 + \lambda \cos^2\left(\alpha_{\beta} - \alpha_{0}\right)\Bigg\}
$$
where $\alpha_{\beta}\in(-\pi,\pi], r_{\beta}\geq 0$ and $\lambda \in \mathbb{R}$. For an orthonormal design matrix, $X^{\T}X = I_{2}$, the normal equations give explicit solutions for the parameter estimates:  
\begin{equation}
\tan 2 \hat\alpha_{\beta, x_0}(\lambda) = \frac{\tilde r_{\beta}^{ 2}\sin 2\tilde\alpha_{\beta} + \lambda \sin2\left(\alpha_{0}\pm\frac{\pi}{2}\right)}{ \tilde r_{\beta}^{ 2}\cos 2\tilde\alpha_{\beta}  + \lambda \cos2\left(\alpha_{0}\pm\frac{\pi}{2}\right)}, \quad 
\hat r_{\beta, x_0}(\lambda) = \tilde r_{\beta}\cos(\tilde\alpha_{\beta} - \hat\alpha_{\beta, x_0}(\lambda)),
\label{eq:tangens}
\end{equation}
The Equation \eqref{eq:tangens} shows that for $\lambda = 0$, the estimated angle and length are equal to the angle and length of the OLS estimate. In the limit $\lambda \to \infty$, the estimated angle converges to either $\hat\alpha_{\beta, x_0}(\lambda) \to \alpha_{0} + \frac{\pi}{2}$, if $\tilde\alpha_{\beta} \in [\alpha_{0},\alpha_{0}+\pi]$, or to $\hat\alpha_{\beta, x_0}(\lambda) \to \alpha_{0} - \frac{\pi}{2}$, if $\tilde\alpha_{\beta} \in [\alpha_{0}-\pi,\alpha_{0}]$, becoming exactly orthogonal to $x_{0}$. The estimated angle and length will hence shrink the prediction for $x_{0}$ towards zero.

In the orthonormal design case, the prediction for $x_{0}$ is in hyperspherical coordinates given by the estimated length and the double tangent expression in Equation \eqref{eq:tangens} 
$$x_{0}^{\T}\hat \beta_{x_0}(\lambda) = r_0 \tilde r_\beta \cos( \alpha_{0} - \tilde \alpha_\beta) \;\Bigg( \frac{1}{2} + \frac{1}{2} \frac{\tilde r_{\beta}^{ 2}- \lambda}{\sqrt{ (\tilde r_{\beta}^2 + \lambda)^{2} - 4 \lambda \tilde r_{\beta}^2 \cos^{ 2}(\alpha_{0}-\tilde\alpha_{\beta})}}\Bigg), %
$$
where $r_0 \tilde r_\beta \cos( \alpha_{0} - \tilde \alpha_\beta) = x_{0}^{\T}\tilde\beta$ is the OLS prediction. The PAN prediction hence equals the OLS prediction multiplied by a shrinkage factor. When $\lambda$ increases, the shrinkage increases and as $\lambda \to \infty$, the factor converges to zero. Importantly, the shrinkage factor depends on 
the angle of the specific covariate vector, such that the shrinkage will \emph{vary} for different $x_{0}$ when $\lambda$ is fixed. The shrinkage term thus explicitly expresses the feature of personalization inherent in the penalty. 



\begin{definition}[Hyperspherical coordinates]\label{def:hyper} 
The Personalized Angle (PAN) estimator in hyperspherical coordinates $\hat\beta_{x_0}=(\hat r_\beta,\hat\alpha_{\beta,1},\dots, \hat\alpha_{\beta,p-1})^{\T}$, for a specific covariate vector $x_{0}$ parametrized by $r_0$  and $\alpha_{0,1},\dots,\alpha_{0,p-1}$ is defined as 
\begin{align*} 
 &\hat\beta_{x_0}(\lambda) = \arg\min 
 \Bigg\{\sum_{i = 1}^n \Big(  y_{i} - r_{\beta} \big(
 \cos(\alpha_{\beta,p-1} -\alpha_{i,p-1})\prod_{j=1}^{p-2}\sin \alpha_{\beta,j} \sin \alpha_{i,j}  \\
& \qquad\qquad\qquad\qquad\qquad\qquad + \sum_{j=2}^{p-2}\cos \alpha_{\beta,j}\cos\alpha_{i,j}\prod_{k=1}^{p-2}\sin \alpha_{\beta,k} \sin \alpha_{i,k} 
\big) \Big)^2 \\
  + & \lambda \Big(\cos(\alpha_{\beta,p-1} -\alpha_{0,p-1})\prod_{j=1}^{p-2}\sin \alpha_{\beta,j} \sin \alpha_{0,j} + \sum_{j=2}^{p-2}\cos \alpha_{\beta,j}\cos\alpha_{0,j} \prod_{k=1}^{p-2}\sin \alpha_{\beta,k} \sin \alpha_{0,k} \Big)^2\Bigg\},
\end{align*}
where $\lambda \in \mathbb{R}$ is a tuning parameter.  
\end{definition}
The hyperspherical parametrization has computational advantages, in particular improved convergence, when obtaining an estimate via a numerical optimizer routine. The subscripts of the regression coefficients and the $L_2$ norm are further suppressed for notational convenience.


\subsection{Orthonormal design case}
Insight regarding the behavior of the PAN penalty in both the Cartesian and hyperspherical coordinates is gained by considering the case of the orthonormal design matrix, $X^TX = I_p$. The PAN estimate and prediction are then given explicitly. 
\begin{lemma}\label{lemma:solution}
Assuming an orthonormal design matrix,  $X^{\T}X= I_{p}$, the length of the PAN estimate is given by
$$\hat r(\lambda) =  \tilde\beta^{\T}\hat\gamma(\lambda) = \left[\frac{1}{2}+\frac{1}{2}\;c(\lambda)\right]^{\frac{1}{2}}\|\tilde\beta\|,$$
while the direction vector of PAN estimate equals the first normalized eigenvector of the $p \times p$ matrix of rank 2
$$M := \tilde\beta\tilde\beta^{\T} - \frac{\lambda_{2}}{\|x_0\|^2} x_{0}x_{0}^{\T},$$ given by
$$ \hat \gamma(\lambda) = \left[\frac{1}{2}+\frac{1}{2}\;c(\lambda)\right]^{\frac{1}{2}}\frac{\tilde\beta}{\|\tilde\beta\|} -\left[\frac{1}{2}-\frac{1}{2}\;c(\lambda)\right]^{\frac{1}{2}}\frac{ \|\tilde\beta\|^{2} x_{0} - (x_{0}^{\T}\tilde\beta) \tilde\beta}{\|\tilde\beta\|\sqrt{\|\tilde\beta\|^{2}\|x_{0}\|^{2} - (x_{0}^{\T}\tilde\beta)^{2}}},$$ 
depending on the tuning parameter, $\lambda$, through
\begin{equation}
c(\lambda) = \frac{\|\tilde\beta\|^{2}(\|\tilde\beta\|^{ 2} + \lambda) -2\lambda(x_{0}^{\T}\tilde\beta)^{ 2}/\|x_{0}\|^{ 2}}{
\|\tilde\beta\|^{2}\sqrt{(\|\tilde\beta\|^{ 2}+\lambda)^{ 2}-4\lambda(x_{0}^{\T}\tilde\beta)^{2}/\|x_{0}\|^{2}}}.
\label{eq:constantPAN}
\end{equation}
The PAN estimate is then given
$$ \hat \beta(\lambda)  = \hat r \;\hat\gamma = \frac{1}{2}\left[1+c(\lambda)\right]\tilde\beta  - \frac{1}{2}\left[1-c^{ 2}(\lambda)\right]^{\frac{1}{2}} \frac{\|\tilde\beta\|^{2} x_{0} - (x_{0}^{\T}\tilde\beta) \tilde\beta}{\sqrt{\|\tilde\beta\|^{2}\|x_{0}\|^{2} - (x_{0}^{\T}\tilde\beta)^{2}}}. $$ 
\end{lemma}
The proof of Lemma \ref{lemma:solution} can be found in the Appendix. For $\lambda = 0$, the constant in Equation \eqref{eq:constantPAN} is $c(0) = 1$, and hence the PAN estimate equals the OLS estimate. In the limit, 
$ \lim_{\lambda\to\infty} c(\lambda) = 1 - \frac{2(x_{0}^{\T}\tilde\beta)^{ 2}}{
\|\tilde\beta\|^{ 2}\|x_{0}\|^{ 2}},$
the length and direction vector converge to 
\begin{align*} \lim_{\lambda\to\infty}\hat r(\lambda) &= \left[1 - \frac{(x_{0}^{\T}\tilde\beta)^{ 2}}{
\|\tilde\beta\|^{ 2}\|x_{0}\|^{ 2}}\right]^{\frac{1}{2}}\|\tilde\beta\|, \\
\lim_{\lambda\to\infty}\hat \gamma(\lambda) &= \frac{\|x_{0}\|}{\sqrt{\|\tilde\beta\|^{ 2}\|x_{0}\|^{ 2}-(x_{0}^{\T}\tilde\beta)^{ 2}}}\left(\tilde\beta-\frac{x_{0}^{ T}\beta}{x_{0}^{ T}x_{0}}x_{0}\right),
\end{align*}
where the direction vector is equal to the normalized projection of $\tilde\beta$ unto $H_{0}$.
From Lemma \eqref{lemma:solution}, it is seen that the PAN estimate depends on the tuning parameter, $\lambda$, through the direction vector. 

\begin{corollary}\label{cor:prediction}
The PAN prediction of the outcome $\hat y_0$ given the covariate vector $x_{0}$ is in Cartesian coordinates given by
\begin{align}
x_{0}^{\T}\hat \beta(\lambda)  =  x_{0}^{\T}\tilde\beta \left[\frac{1}{2}+\frac{1}{2}\frac{\|\tilde\beta\|^{2}-\lambda}{ \sqrt{(\|\tilde\beta\|^{2}+\lambda)^{2} - 4\lambda(x_{0}^{\T}\tilde\beta)^{2}/\|x_{0}\|^{2}}}\right],
\end{align}
where $x_{0}^{\T}\tilde\beta$ is the OLS prediction.
\end{corollary}
The proof of Corollary \ref{cor:prediction} can be found in the Appendix. 
In the limit, $\lambda \to \infty$, the prediction converges to $x_{0}^{\T}\hat \beta(\lambda)  \to  x_{0}^{\T}\tilde\beta \left[1/2-1/2\right] = 0$,
while for $\lambda \to -\infty$, the prediction converges to $x_{0}^{\T}\hat \beta(\lambda)  \to  x_{0}^{\T}\tilde\beta \left[1/2+1/2\right] = x_{0}^{\T}\tilde\beta$, the OLS prediction. 
In the latter case where $\lambda$ decreases from 0, the prediction will first increase or expand. At a certain value of $\lambda$, however, the length of the regression vector will cancel out the effect of the rotation in the direction vector, such that the prediction decreases and converges to the OLS prediction.


\section{Simultaneous penalization of length and angle}
The PAN penalty can be combined with the ridge penalty with the resulting PAN-ridge estimate depending on two tuning parameters
\begin{align}
\hat{\beta}(\lambda_{1},\lambda_{2}) = \arg\min_{\beta}\left\{\sum_{i = 1}^n \left(  y_{i} - x_{i}^{\T}\beta \right)^2 + \lambda_{1}\beta^{\T}\beta + \frac{\lambda_{2}}{x_{0}^{\T}x_{0}}\frac{\beta^{\T}x_{0}x_{0}^{\T}\beta}{\beta^{\T}\beta}\right\}, 
\label{eq:penalized2}
\end{align}
where $\lambda_{1} \geq 0$ and $\lambda_{2} \in \mathbb{R}$. As with PAN regression, a hyperspherical parametrization of the objective function gives computational advantages. 

For an orthonormal design matrix, the length of the PAN-ridge estimate has the same form as the length of the PAN estimate, but with the ridge estimate taking the role of the OLS:
\begin{align*} 
\hat r(\lambda_{1},\lambda_{2}) = \tilde\beta(\lambda_{1})^{\T} \hat\gamma(\lambda_{1},\lambda_{2}), 
\end{align*}
where $\tilde\beta(\lambda_{1})$ only depends on the first tuning parameter, $\lambda_{1}$. Similarly, the direction vector equals the normalized eigenvector of the following $p \times p$ matrix of rank 2 
$$M := (1+\lambda_{1})\tilde\beta(\lambda_{1})\tilde\beta(\lambda_{1})^{\T} - \lambda_{2} \gamma_{0}\gamma_{0}^{\T}.$$ 
The PAN-ridge regression coefficient vector is then given by
$$ \hat \beta(\lambda_{1},\lambda_{2}) = \frac{1}{2(1+\lambda_{1})}\left[1+C(\lambda_{1},\lambda_{2})\right]\tilde\beta  - \frac{1}{2(1+\lambda_{1})}\left[1-C(\lambda_{1},\lambda_{2})^{ 2}\right]^{\frac{1}{2}} \frac{\|\tilde\beta\|^{2} x_{0} - (x_{0}^{\T}\tilde\beta) \tilde\beta}{\sqrt{\|\tilde\beta\|^{2}\|x_{0}\|^{2} - (x_{0}^{\T}\tilde\beta)^{2}}}, $$ 
where 
\begin{equation}
C(\lambda_{1},\lambda_{2}) = \frac{\|\tilde\beta\|^{2}(\|\tilde\beta\|^{ 2} + (1+\lambda_{1})\lambda_{2}) -2(1+\lambda_{1})\lambda_{2}(x_{0}^{\T}\tilde\beta)^{ 2}/\|x_{0}\|^{ 2}}{
\|\tilde\beta\|^{2}\sqrt{(\|\tilde\beta\|^{ 2}+(1+\lambda_{1})\lambda_{2})^{ 2}-4(1+\lambda_{1})\lambda_{2}(x_{0}^{\T}\tilde\beta)^{2}/\|x_{0}\|^{ 2}}}.
\label{eq:constant}
\end{equation}
For $\lambda_{1}=0$, the constant in Equation \eqref{eq:constant} equals the constant in Equation \eqref{eq:constantPAN}: $C(0,\lambda_{2}) =c(\lambda_{2})$, while for $\lambda_{2} = 0$, $C(\lambda_{1},0) = 1$ and the estimate reduces to standard ridge regression in the orthonormal case, $\hat \beta(\lambda_{1},0) = \tilde \beta/(1+\lambda_{1})$. The PAN-ridge prediction for a specific covariate $x_{0}$ is consequently 
\begin{align}
\hspace{-10pt}x_{0}^{\T}\hat \beta(\lambda_{1},\lambda_{2})  =  \frac{x_{0}^{\T}\tilde\beta }{1+\lambda_{1}}\left[\frac{1}{2}+\frac{1}{2}\frac{\|\tilde\beta\|^{2}-(1+\lambda_{1})\lambda_{2}}{ \sqrt{(\|\tilde\beta\|^{2}+(1+\lambda_{1})\lambda_{2})^{2} - 4(1+\lambda_{1})\lambda_{2}(x_{0}^{\T}\tilde\beta)^{2}/\|x_{0}\|^{2}}}\right]. \label{eq:PANridgeprediction}
\end{align}
It is seen that both the PAN-ridge estimate and the prediction depend on both tuning parameters, $\lambda_{1}$ and $\lambda_{2}$, in particular through the direction vector. In the high-dimensional case ($p > n$) or if $X^{\T}X$ is not of full rank, the PAN-ridge estimate can be expressed only in terms of the ridge estimate. 

\begin{figure}%
\centering\includegraphics[width=0.65\columnwidth]{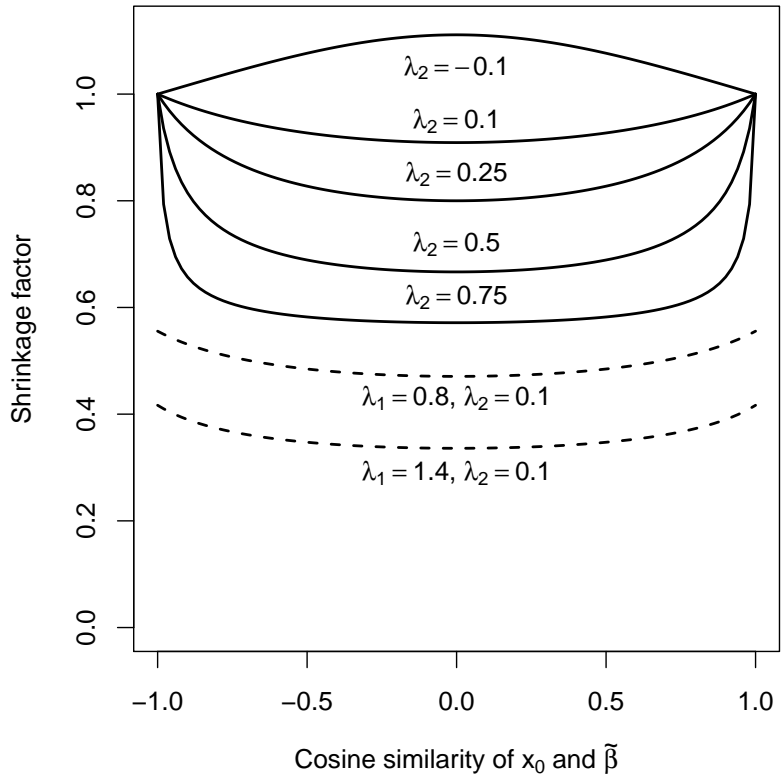}
\caption{The shrinkage factor of the PAN-ridge prediction in Equation \eqref{eq:PANridgeprediction} as a function of the cosine similarity $\frac{x_{0}^{\T}\tilde\beta}{\|\tilde\beta\|\|x_0\|}$ for fixed $\|\tilde\beta\|= \|x_0\|=1$ and different values of $\lambda_1$ and $\lambda_2$. The shrinkage factors of the PAN penalty (with $\lambda_1 = 0$) are shown by solid lines, and the factors of the combined PAN-ridge method are shown by dashed lines.}%
\label{fig:shrinkage}
\end{figure}

Figure \ref{fig:shrinkage} shows the shrinkage factor of the PAN-ridge prediction in Equation \eqref{eq:PANridgeprediction} as a function of the normalized inner product or cosine similarity 
$$ \text{CosSim}(x_0,\tilde\beta)= \frac{x_{0}^{\T}\tilde\beta}{\|\tilde\beta\|\|x_0\|},$$ 
for fixed lengths, $\|\tilde\beta\| = \|x_0\| = 1$. 
Different levels of PAN penalization (with $\lambda_1$ set to zero) are shown by solid lines, while the combined PAN-ridge shrinkage with the values of $\lambda_1$ and $\lambda_2$ are shown by dashed lines. The shrinkage factor for a positive PAN parameter is strongest for the cosine similarity values closest to zero and increases to 1 when the cosine similarity approaches 1 and -1. The shrinkage becomes stronger with an increasing PAN parameter, but inverts if the parameter becomes negative. Then the ``expansion'' factor is strongest for the smallest cosine similarities in absolute value. The ridge parameter, on the other hand, controls the overall level of penalization and shifts the level of the shrinkage curve downwards with increasing values.  

\subsection{Prediction error}
The main aim of personalizing a prediction is to lower the prediction error for each individual covariate vector, $x_{0}$, instead of minimizing the average prediction error \citep{hellton2018fridge,huang2019tuning}. The predictive performance of the regression methods can be evaluated by the mean squared error (MSE) of the prediction for covariate vector $x_0$ under the linear model
$$ \mse(x_{0},\beta, \lambda_{1},\lambda_{2})  = E\left[(x_{0}^{\T} \hat\beta(\lambda_{1},\lambda_{2}) - x_{0}^{\T}\beta)^2 \mid X\right], $$
related to the prediction error as $ E\left[(x_{0}^{\T} \hat\beta(\lambda_{1},\lambda_{2}) - y_0)^2 \mid X\right] =  \mse(x_{0},\beta, \lambda_{1},\lambda_{2}) + \sigma^2.$
For a given $x_0$, we will compare the predictions in terms of the MSE to omit the intrinsic error $\sigma^2$. Later, the average MSE is used to evaluate the prediction performance over a given sample. We first present a lemma demonstrating the behavior of the optimal $\lambda_{2}$ in terms of minimum MSE. A scaling of the design matrix is introduced to ensure the asymptotic convergence of $\tilde\beta$. 

\begin{lemma}\label{lem:derivative}
Under a scaled orthogonal design matrix, $X^\T X = n I_{p}$, the derivative of the mean square error with respect to $\lambda_{2}$ evaluated at $0$ is given by 
\begin{align*}
\left.\frac{\partial\mse(x_{0},\beta, \lambda_{1},\lambda_{2})}{\partial\lambda_{2}}\right|_{\lambda_{2}=0} & = C_1 \bigg( \lambda_{1}(x_{0}^{\T}\beta)^{2} -  \sigma^{ 2}\|x_{0}\|^{2} \Big(1-4 \frac{(x_{0}^{\T}\beta)^{2}}{\|x_{0}\|^{2}\|\beta\|^{2}}\Big) \bigg) + O\left(\frac{1}{n^{3}}\right), 
\end{align*}
with the positive constant $C_1 =  \frac{2}{n}\frac{(\|x_{0}\|^{2}\|\beta\|^{2}-(x_{0}^{\T}\beta)^{2})}{(n+ \lambda_{1})\|x_{0}\|^{2}\|\beta\|^{ 4}}$.
\end{lemma}
The proof of Lemma \ref{lem:derivative} can be found in the Appendix. 
As the value $\lambda_{2}= 0$ of the PAN-ridge estimator corresponds to ridge regression, Lemma \ref{lem:derivative} shows when the PAN penalty improves the mean squared prediction error compared to ridge and OLS. We prove the results first for the PAN estimator ($\lambda_{1}=0$) and then for the combined PAN-ridge estimator. 

\begin{theorem}\label{thm:uniform}
Let $\lambda_{1} = 0$ and assume an orthogonal design matrix, $X^\T X = n I_{p}$. 
Then if $|x_{0}^{\T}\beta| < \frac{1}{2}\|x_{0}\||\beta\|$, there exists a $\lambda_{2} > 0$, 
and if $|x_{0}^{\T}\beta| > \frac{1}{2}\|x_{0}\||\beta\|$, there exists a $\lambda_{2} < 0$, for which the mean squared error asymptotically as $n\to\infty$ satisfies the inequality 
$$\mse(x_{0},\beta, 0,\lambda_{2}) < \mse(x_{0},\beta, 0,0) = \mse_{OLS}(x_{0},\beta).$$
When $|x_{0}^{\T}\beta| = \frac{1}{2}\|x_{0}\||\beta\|$, the minimum of the mean squared error is asymptotically obtained for $\lambda_{2} = 0$. 
\end{theorem}
For the PAN estimator, Theorem \ref{thm:uniform} demonstrates that the sign of the optimal value for $\lambda_{2}$ is dependent on whether the absolute value of $x_{0}^{\T}\beta$ is smaller or larger than $\frac{1}{2}\|x_{0}\||\beta\|$. This corresponds to the absolute value of cosine similarity between $x_{0}$ and $\beta$ being smaller or larger than 0.5. For small cosine similarities, the optimal PAN tuning parameter is hence positive, while for large cosine similarities the optimal value will be negative. This result can be extended to include the ridge penalty. 
\begin{theorem}\label{thm:uniform2}
Assume an orthogonal design matrix, $X^\T X = n I_{p}$. For $\lambda_{1} > 0$, if $\lambda_{1} < \lambda_{1}^{*}$, there exists a $\lambda_{2} > 0$ for which the mean squared error asymptotically satisfies 
 $$\mse(x_{0},\beta, \lambda_{1},\lambda_{2}) < \mse(x_{0},\beta, \lambda_{1},0) = \mse_{ridge}(x_{0},\beta,\lambda_{1}),$$
while if $ \lambda_{1} > \lambda_{1}^{*}$, there exists a $\lambda_{2} < 0$ for which the mean squared error asymptotically satisfies 
$$\mse(x_{0},\beta, \lambda_{1},\lambda_{2}) < \mse(x_{0},\beta, \lambda_{1},0) = \mse_{ridge}(x_{0},\beta,\lambda_{1}),$$
where 
$$\lambda_{1}^{*} =  \frac{\sigma^{ 2}\|x_{0}\|^{2}}{(x_{0}^{\T}\beta)^{ 2}} 
\Big(1-4 \frac{(x_{0}^{\T}\beta)^{2}}{\|x_{0}\|^{2}\|\beta\|^{2}}\Big)
.$$
When $\lambda_{1} = \lambda_{1}^{*}$, the minimum of the mean squared error is asymptotically obtained for $\lambda_{2} = 0$. 
\end{theorem}
The proof of Theorem \ref{thm:uniform} and \ref{thm:uniform2} can be found in the Appendix. 
Based on Theorem \ref{thm:uniform2}, the PAN-ridge estimate will always have smaller mean squared prediction error than ridge regression asymptotically, except when the ridge tuning parameter is exactly equal to $\lambda^{*}_{1}$. This value is related to the tuning parameter value minimizing the prediction risk of $x_0$, the oracle focused ridge tuning parameter \citep{hellton2018fridge}:
$$\lambda_{x_0} = \frac{\sigma^2\|x_{0}\|^{2}}{(x_{0}^{\T}\beta)^2},$$
When $\lambda_{1} < \lambda_{1}^{*}$, the level of penalization can be viewed as being too small compared to the optimal level, and a positive PAN tuning parameter $\lambda_{2}>0$ can be used to introduce additional penalization further shrinking the prediction. On the other hand, if the ridge tuning parameter is larger than the optimal value, $\lambda_{1} > \lambda_{1}^{*}$, the level of penalization can be viewed as being too strong, such that the ridge prediction is shrunken too much towards zero. Allowing the PAN tuning parameter to be negative, $\lambda_{2}<0$, essentially expands the prediction away from zero, reducing the shrinkage and adjusting the overall level of penalization closer to the optimal value. 

\begin{remark}
Interestingly, Lemma \ref{thm:uniform2} also reveals that the benefit of estimating a common PAN tuning parameter for an entire sample may depend on the dimension. If the covariates are assumed to be standard normally distributed in $p$ dimensions, $x_0 \sim N(0,I_p)$, for a fixed, arbitrary, $\beta$, the normalized inner product, $z = x_0^\T\beta/(\|x_0\|\|\beta\|)$, follows the distribution 
$$f_p(z) = \frac{1}{\sqrt{\pi}}\frac{\Gamma(\frac{p}{2})}{\Gamma(\frac{p-1}{2})}\left(1-z^2\right)^{\frac{p-3}{2}}, \quad \text{ for } -1 < z < 1, $$
after \citet{cho2009inner}, where $\Gamma(\cdot)$ is the gamma function. The proportion of observations with normalized inner product between -1/2 and 1/2, i.e.\ they benefit from a positive PAN parameter, will greatly increase with the dimension. For $p=2$ and $=3$, this proportion is $1/3$ and $1/2$ respectively. Hence, in dimension two and three, around half of the observations will benefit from a negative tuning parameter value, while the other half will benefit from a positive value. Selecting a single, common, tuning parameter, either positive or negative, will therefore be unsuitable for half of the data. 
However, as the proportion requiring a positive PAN value increases rapidly with $p$, to 74.7\% for $p=6$, 95.1\% for $p= 15$ and 99.6\% for $p=30$, estimating a common tuning parameter value will be more beneficial in higher dimension. 
\end{remark}

\section{Simulation}
In this section, we present a simulation study comparing PAN and PAN-ridge regression to OLS and ridge regression. We simulated 200 data sets consisting of 50 observations from a linear model with $6$ and $15$ variables:
$$ y_i = x_i^{\T} \beta + \varepsilon_i,$$
where the noise is normally distributed $\varepsilon_i \sim N(0,\sigma^2)$ with $\sigma = 3$. The data matrix was simulated from a standard normal distribution and scaled to be orthonormal, such that $X^{\T} X = I_p$. An independent test set with 1000 observations was predicted for each simulation. To select the tuning parameter, we used the parametric bootstrap procedure described in Section \ref{sec:tuning} as an alternative to cross-validation. 

\begin{table}
\caption{\label{tab:simulation} The mean squared error over 200 simulations with $n = 50$ and $\sigma = 3$. The simulations were carried out for $p=6,15$ and four values of equal regression coefficients.} 
\begin{tabular}{lccccc}
 &  \multicolumn{4}{c}{$p=6$} \\
Method & $\beta_j = 0.05$ & $\beta_j = 0.10$ & $\beta_j = 0.15$ &$\beta_j = 0.20$\\
 \hline 
  OLS & 0.124 & 0.126 & 0.123 & 0.124 \\ 
  PAN & 0.037 & 0.060 & 0.089 & 0.104 \\ 
  Ridge & 0.048 & 0.062 & 0.082 & 0.095 \\ 
  PAN-ridge (fixed $\lambda_1$, oracle) & 0.014 & 0.043 & 0.070 & 0.087 \\ 
  PAN-ridge (fixed $\lambda_1$, estimated) & 0.044 & 0.066 & 0.092 & 0.101 \\ 
  PAN-ridge ($\lambda_1,\lambda_2$) & 0.039 & 0.061 & 0.088 & 0.101 \\    \hline
  &  \multicolumn{4}{c}{$p=15$} \\
  Method & $\beta_j = 0.05$ & $\beta_j = 0.10$ & $\beta_j = 0.15$ &$\beta_j = 0.20$\\ \hline
OLS & 0.302 & 0.299 & 0.297 & 0.297 \\ 
  PAN & 0.067 & 0.124 & 0.171 & 0.214 \\ 
  Ridge & 0.102 & 0.141 & 0.177 & 0.211 \\ 
  PAN-ridge (fixed $\lambda_1$, oracle) & 0.034 & 0.101 & 0.155 & 0.198 \\ 
  PAN-ridge (fixed $\lambda_1$, estimated) & 0.066 & 0.129 & 0.175 & 0.215 \\ 
  PAN-ridge  ($\lambda_1,\lambda_2$) & 0.097 & 0.153 & 0.189 & 0.219 
  \end{tabular}
\end{table}

Table \ref{tab:simulation} shows the average MSE over the test set sample, $\frac{1}{1000}\sum_{i=1}^{1000}(x_i^{\T}\hat\beta -x_i^{\T}\beta)^2$,
for the OLS, ridge and PAN regression estimates in addition to the PAN-ridge combination, averaged over 200 simulations. The simulations were performed for four scenarios of increasing signal strengths with equal regression coefficients: 1) $\beta_j = 0.05, \forall j$, 2) $\beta_j = 0.1, \forall j$, 3) $\beta_j = 0.15, \forall j$ and 4) $\beta_j = 0.2, \forall j$. The four values of $\beta_j$ were chosen such that ridge regression would yield an improvement compared to OLS. The tuning parameters for ridge and PAN regression were found using the parametric bootstrap procedure with the other tuning parameter fixed to 0. For the PAN-ridge combination, the tuning parameters were selected following two different strategies: first, $\lambda_1$ was fixed to the optimal value found for ridge regression, selecting only $\lambda_2$, and second, both $\lambda_1$ and $\lambda_2$ were selected simultaneously. When selecting $\lambda_2$ with the fixed ridge tuning parameter, we used the parametric bootstrap procedure based both on the OLS estimates, $\tilde\beta, \hat\sigma^2$, and the true parameter values $\beta, \sigma^2$, referred to as the oracle tuning. The oracle tuning parameter is the value of $\lambda_2$ which would be optimal if $\beta$ and $\sigma^2$ were in fact known. For all instances, the number of bootstrap samples was set to $B=2000$. 

The results of Table \ref{tab:simulation} show that for $p=6$, PAN regression improves on ridge regression for $\beta_j=0.05$ and $0.10$, the smallest signal strengths, while ridge regression is better for $\beta_j=0.15$ and $0.20$. For $p=15$, PAN performs better than ridge regression for $\beta_j = 0.05$, $0.10$ and $0.15$, showing the effect of the dimension.  

For the PAN-ridge combination, it is seen that for the fixed ridge parameter, the oracle PAN tuning always gives a lower prediction error compared to ridge regression as supported by Theorem \ref{thm:uniform2}. When the PAN tuning parameter is estimated, however, the PAN-ridge combination improves on ridge regression for $\beta_j = 0.05$, $0.10$ and $0.15$ when $p=15$, but only for $\beta_j=0.05$ when $p=6$. This suggests that the PAN tuning parameter is difficult to estimate correctly, in particular if the dimension is small. When both tuning parameters are selected simultaneously, the MSE is only clearly lower compared to ridge regression when $\beta_j=0.05$ for both $p=6$ and 15. 

\subsection{Selecting the tuning parameter}\label{sec:tuning}
We propose to select the tuning parameter in PAN and PAN-ridge regression by the following procedure based on parametric bootstrap \citep{efron1994introduction}:
\begin{enumerate}
 \item Use the OLS estimates $\tilde \beta$ and  $\tilde \sigma^2$ as plug-in estimates to simulate $r=1,\dots,B$ bootstrap samples of $n$ observations $Y^{(r)} = [y_{1}^{(r)}, \dots, y_{n}^{(r)}]^{\T}$ from 
 $$ y_{i}^{(r)} = x_{i}^{\T}\tilde\beta + \varepsilon_i, \quad \varepsilon_i \sim N(0,\hat \sigma^2), \qquad i = 1, \dots, n.$$
 \item Over a suitable grid of $\lambda_1$ and $\lambda_2$, hold the tuning parameter values fixed: 
 \begin{itemize}
  \item calculate $x_i^{\T}\hat\beta_{\lambda_1,\lambda_2}^{(r)}$ for each bootstrap sample and $x_i, i =1, \dots, n$,
  \item average the mean squared error $(x_{i}^{\T}\hat\beta_{\lambda_1,\lambda_2}^{(r)} - x_{i}^{\T}\tilde\beta)^2$ over all $i$ and $r$. 
 \end{itemize}
 \item Select the tuning parameter values, $\hat \lambda_1$ and $\hat \lambda_2$, with the smallest mean squared error over the grid of $\lambda_1$ and $\lambda_2$.   
\end{enumerate}
The procedure was also used by \citet{hellton2018fridge} to estimate the personalized tuning parameter in ridge regression.  

\section{Example: Prostate cancer data}
We demonstrate PAN regression and the PAN-ridge combination on a classical dataset previously used to illustrate penalized regression methods \citep{tibshirani1996regression}. The dataset examines the relation between prostate specific antigen (PSA) and clinical measurements in 97 prostate cancer patients \citep{stamey1989prostate}. We predict the log PSA values based on the six covariates; log tumor volume (\verb|lcavol|), log tumor weight (\verb|lweight|), age  (\verb|age|), log of benign prostatic hyperplasia amount (\verb|lbph|), seminal vesicle invasion and log of capsular penetration (\verb|lcp|). The tuning parameters for the ridge and PAN penalties, $\lambda_1$ and $\lambda_2$, are both determined following the procedure described in Section \ref{sec:tuning}. We further estimate the out-of-sample prediction error by leave-one-out cross-validation. 

\begin{figure}%
\centering\includegraphics[width=\columnwidth]{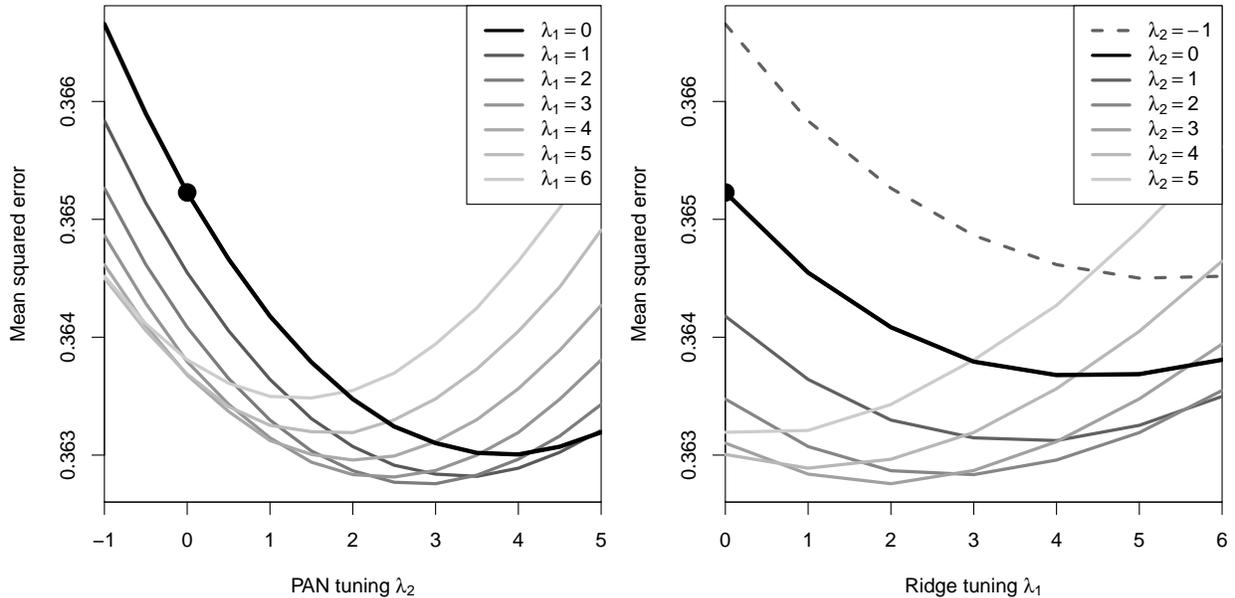}
\caption{The parametric bootstrap prediction error as a function of $\lambda_{2}$ for different values of $\lambda_{1}$. Standard ridge regression prediction error corresponds to $\lambda_{2}=0$ (thick black line) and the OLS prediction error corresponds to $\lambda_{1}=0$ and $\lambda_{2}=0$ (circle).}%
\label{prostate_tuning}%
\end{figure}

First, we find each of the optimal PAN and ridge tuning parameters by fixing the other tuning value to zero in the bootstrap procedure, and then both tuning parameters are optimized simultaneously in a 2-dimensional grid. Figure \ref{prostate_tuning} shows the average MSE from the parametric bootstrap procedure as a function of $\lambda_{2}$ for different values of $\lambda_{1}$, and reversely. From the left panel of Figure \ref{prostate_tuning}, we find the optimal PAN tuning parameter to be $\lambda_2 = 2.5$ and from the right panel, the optimal ridge tuning parameter to be $\lambda_{1} = 4$ (as the minima of the thick black lines). For all shown values of $\lambda_{1}$, in the left panel, the minimum of the MSE is found for a positive PAN tuning parameter. It is also seen that as the ridge tuning parameter increases, the value of $\lambda_{2}$ yielding the minimum decreases. This occurs as a stronger level of ridge penalization must be counteracted by a smaller, and possibly negative, PAN tuning. Based on Figure \ref{prostate_tuning}, we find that the simultaneously optimal tuning parameter values are $\lambda_1 = 2$ and $\lambda_{2}= 3$. 

\begin{table}
\caption{\label{tab:PANregression} The prediction and regression coefficients of OLS and PAN for the observations with the four largest and smallest cosine similarities between $x_0$ and $\tilde \beta$ in absolute value.} \vspace{3mm}
\centering 
\begin{tabular}{rr|rrrrrrrrr}
  \hline
\multicolumn{2}{l|}{Observation} & 61 & 60 & 46 & 30 && 92 & 23 & 2 & 4 \\ 
  \hline
\multicolumn{2}{l|}{Cosine similarity}  
& -0.010 & 0.016 & 0.022 & 0.023  && 0.760 & -0.766 & -0.778 & -0.796  \\ 
  \multicolumn{2}{l|}{OLS prediction} 
  & -0.021 & 0.016 & 0.020 & 0.033  && 1.328 & -1.170 & -1.457 & -1.578  \\ 
  \multicolumn{2}{l|}{PAN prediction}
  & -0.020 & 0.014 & 0.016 & 0.027  && 1.028 & -1.147 & -1.426 & -1.549  \\ 
  &&&&&&&& \\
  &OLS& \multicolumn{4}{|c}{PAN coefficients} && \multicolumn{4}{c}{PAN  coefficients} \\
  \verb|lcavol| & 0.578 & 0.578 & 0.578 & 0.576 & 0.578&  & 0.572 & 0.595 & 0.589 & 0.587  \\ 
   \verb|lweight| & 0.216 & 0.217 & 0.216 & 0.217 & 0.217 & & 0.243 & 0.222 & 0.226 & 0.225  \\ 
   \verb|age| & -0.107 & -0.107 & -0.107 & -0.107 & -0.108 &&  -0.098 & -0.135 & -0.134 & -0.134  \\ 
   \verb+lbph+ & 0.130 & 0.130 & 0.129 & 0.128 & 0.131 && 0.091 & 0.130 & 0.121 & 0.124  \\ 
   \verb+svi+ & 0.279 & 0.279 & 0.278 & 0.279 & 0.281  && 0.180 & 0.321 & 0.322 & 0.325  \\ 
   \verb+lcp+ & -0.050 & -0.050 & -0.049 & -0.048 & -0.052 && 0.037 & -0.117 & -0.111 & -0.112 
\end{tabular}
\end{table}

Table \ref{tab:PANregression} displays the personalized PAN regression coefficients (with $\lambda_2 = 2.5$) calculated for the four observations, or patients, with the smallest and the four patients with the largest cosine similarity between $x_0$ and $\tilde \beta$ in absolute value. This shows that even though the observations with the smallest cosine similarity experience the largest shrinkage factor (as seen in Figure \ref{fig:shrinkage}), the observations with the highest cosine similarity experience the largest change in the regression coefficients compared to the OLS coefficients. The personalized regression coefficients of observation 61 barely change, while for observation 4, the parameter of \verb|svi| changes from 0.279 to 0.325 and the parameter of \verb|lcp| from -0.050 to -0.112. The leave-one-out prediction error is shown in Table \ref{testerror}. The PAN-ridge method has the lowest error, even though the prediction performance of the different methods is very similar. The estimation of the tuning parameter should also be included in the cross-validation step, but as only a single observation is removed for each fold, the change in tuning parameter will be smaller than the bootstrap simulation error and the approximation error caused by the grid of tuning parameters. 

\begin{table}
\caption{\label{testerror}The leave-one-out cross-validation prediction error of the OLS, PAN regression, ridge and PAN-ridge regression.} \vspace{3mm}
\centering 
\begin{tabular}{lcccc}
  \hline
Method $(\lambda_{1}, \lambda_{2})$ & OLS (0,0)  & PAN (0,2.5) & Ridge (4,0) & PAN-ridge (2,3) \\
  \hline
Test error &  0.3903 & 0.3875 & 0.3887 & 0.3874\\
\end{tabular}
\end{table}

\section{Discussion}
We have introduced an inherently personalized regression penalty, constructed to produce individualized regression coefficients and predictions.  The PAN penalty has the advantage over other personalized prediction approaches \citep{hellton2018fridge,huang2019tuning} that a common tuning parameter can be chosen overall based on a training set. 
The PAN penalty can be defined in both Cartesian and hyperspherical coordinates. The Cartesian formulation (Definition \ref{def:cartesian}) enables simple exact expressions in the orthonormal case, while the hyperspherical formulation (Definition \ref{def:hyper}) yields a more computationally efficient objective function. 
The current formulation of the penalty requires additional norm regularization, e.g.\ the $L_{2}$ norm, to be applicable in a high-dimensional setting ($p\gg n$). This is demonstrated by the PAN-ridge combination, where the ridge estimate takes the role of the OLS estimate in PAN regression. Efficient estimation algorithms for the high-dimensional setting need to be developed. 

Due to the structure of the PAN penalty the tuning parameter may be both positive and negative, in  contrast to other penalization methods. This introduces challenges when selecting the tuning parameter value. Initial investigation revealed that (leave-one-out) cross-validation did not work well for PAN, as high variability could obscure the sign the tuning parameter when the optimal value is close to zero. A more stable procedure, such as a parametric bootstrap approach, was proposed instead, yielding good results in simulations. However, as the procedure depends on a plug-in estimate, extensions to higher dimension require further work. Alternative procedures, e.g.\ marginal maximum likelihood or a Bayesian framework, should also be studied. 

PAN regression also has a Bayesian formulation which may be beneficial, for instance, for selecting the tuning parameter. Here the penalty corresponds to a Bayesian prior following the generalized von Mises distribution \citep{gatto2007generalized}. Future work includes to explore other versions of the PAN penalty, i.e.\ corresponding to the lasso or $L_1$ norm in the angle space. Further, the PAN penalty can be extended to logistic regression and generalized linear models and to other methods requiring regularization such as smoothing spline regression or graphical models. 



\appendix

\section{Appendix}

\subsection{Proof of Lemma~\ref{lemma:solution}}
Suppose $X$ is an $n \times p$ matrix of full rank. The gradient of the penalized residual sum-of-squares (penRSS) in Equation \eqref{penRSS} is
\begin{equation}
\frac{\partial \text{penRSS}}{\partial \beta} = - 2X^{\T}Y + 2X^{\T}X \beta + 2\lambda' \frac{x_{0}x_{0}^{\T}}{\beta^{\T}\beta}\beta - 2\lambda' \frac{(x_{0}^{\T}\beta)^{2}}{(\beta^{T }\beta)^{2}}\beta,
\label{eq:derivative}
\end{equation}
where $\lambda'  = \lambda/\|x_{0}\|^{ 2}$. Assume an orthonormal design matrix $X^{\T}X = I_{p}$. By setting the gradient to 0 and multiplying by $\beta^{ T}$ from the left, the last terms cancel such that $\hat\beta^{\T}\hat\beta =  Y^{\T}X\hat\beta = \tilde\beta^{\T}\hat\beta = \sqrt{\hat\beta^{\T}\hat\beta}\tilde\beta^{\T}\hat\gamma$, hence 
$$ \hat r_{\beta} = \tilde\beta^{\T}\hat\gamma.$$
By factoring out and multiplying \eqref{eq:derivative} by $r_{\beta}$, the estimated direction vector fulfills
$$ - \tilde\beta\tilde\beta^{\T}\hat\gamma + (\hat\gamma^{\T}\tilde\beta\tilde\beta^{\T}\hat\gamma)\hat\gamma + \lambda' x_{0}x_{0}^{\T}\gamma + \lambda'(\hat\gamma^{\T}x_{0}x_{0}^{\T}\hat\gamma) \hat\gamma = 0,
$$
such that 
$ (\hat\gamma^{\T}M\hat\gamma)\;\hat\gamma  = M\hat\gamma$, where $M :=\tilde\beta\tilde\beta^{\T} - \lambda' x_{0}x_{0}^{\T}.$
As $\hat\gamma^{\T}M\hat\gamma$ is scalar, $\hat\gamma$ will be equal to a normalized eigenvector of $M$. 

For linearly independent $\tilde\beta$ and $x_{0}$, and $\lambda\neq0$, the rank of $M$ is 2. The range of $M$ is spanned by the orthonormal vectors
\begin{equation}
u_{1} = \frac{\tilde\beta}{\|\tilde\beta\|}, \quad u_{2} = \frac{ \|\tilde\beta\|^{2} x_{0} - (x_{0}^{\T}\tilde\beta) \tilde\beta}{\|\tilde\beta\|\sqrt{\|\tilde\beta\|^{2}\|x_{0}\|^{2} - (x_{0}^{\T}\tilde\beta)^{2}}}.
\label{eq:basis}
\end{equation}
Hence the normalized eigenvectors of $M$ are equal to $(u_{1},u_{2})\eta$ where $\eta$ are the normalized eigenvectors of the $2 \times 2$ matrix, $\tilde M$, for any $p$: 
$$ \tilde M =  \begin{bmatrix} \|\tilde\beta\|^{ 2} - \lambda \frac{(x_{0}^{ T}\tilde\beta) ^{2}}{\|\tilde\beta\|^{ 2}\|x_{0}\|^{ 2}} &  - \lambda \frac{x_{0}^{ T}\tilde\beta}{\|x_{0}\|^{ 2}\|\tilde\beta\|}\sqrt{\|x_{0}\|^{ 2}-\frac{(x_{0}^{ T}\tilde\beta) ^{2}}{\|\tilde\beta\|^{ 2}}} \\
 - \lambda \frac{x_{0}^{ T}\tilde\beta}{\|\tilde\beta\|\|x_{0}\|^{ 2}}\sqrt{\|x_{0}\|^{ 2}-\frac{(x_{0}^{ T}\tilde\beta) ^{2}}{\|\tilde\beta\|^{ 2}}} & - \frac{\lambda}{\|x_{0}\|^{ 2}} \left(\|x_{0}\|^{ 2} -\frac{(x_{0}^{ T}\tilde\beta) ^{2}}{\|\tilde\beta\|^{ 2}}\right) 
\end{bmatrix}. 
$$
The two eigenvectors with positive and negative sign give four stationary points for the penalized RSS in Equation \eqref{penRSS}. For the choice of basis in Equation \eqref{eq:basis}, the global minimum is given by the first eigenvector of $\tilde M$ with a positive first entry. For a matrix, $\begin{bmatrix} \;a & -c \\ -c & \;b \end{bmatrix}, c>0$, this eigenvector is given as
$$\eta_{1} = \left( \left[\frac{1}{2} + \frac{a-b}{2\sqrt{(a-b)^{2}+4c^{2}}}\right]^{\frac{1}{2}}, - \left[\frac{1}{2} - \frac{a-b}{2\sqrt{(a-b)^{2}+4c^{2}}}\right]^{\frac{1}{2}}\right)^{ T},$$ 
such that the direction vector is 
$$\hat\gamma = \left[\frac{1}{2}+\frac{1}{2}\;c(\lambda)\right]^{\frac{1}{2}} u_{1} - \left[\frac{1}{2}-\frac{1}{2}\;c(\lambda)\right]^{\frac{1}{2}} u_{2}, \quad c(\lambda) = \frac{\|\tilde\beta\|^{2}(\|\tilde\beta\|^{ 2} + \lambda) -2\lambda(x_{0}^{\T}\tilde\beta)^{ 2}/\|x_{0}\|^{ 2}}{
\|\tilde\beta\|^{2}\sqrt{(\|\tilde\beta\|^{ 2}+\lambda)^{ 2}-4\lambda(x_{0}^{\T}\tilde\beta)^{2}/\|x_{0}\|^{ 2}}}.$$
As the vector $u_{2}$ is orthogonal to $\tilde \beta$, the length of the PAN estimate is
$$\hat r_{\beta}=  \tilde\beta^{\T}\hat\gamma = \left[\frac{1}{2}+\frac{1}{2}\;c(\lambda)\right]^{\frac{1}{2}}\|\tilde\beta\|.$$

\subsection{Proof of Corollary~\ref{cor:prediction}}
The prediction for $x_0$ is given 
\begin{align*}
x_{0}^{\T}\hat \beta(\lambda) & = \hat r_{\beta} x_{0}^{\T} \hat\gamma
= \frac{1}{2}\left[1+c(\lambda)\right]x_{0}^{\T}\tilde\beta  - \frac{1}{2}\left[1-c^{ 2}(\lambda)\right]^{\frac{1}{2}} \sqrt{\|\tilde\beta\|^{2}\|x_{0}\|^{2} - (x_{0}^{\T}\tilde\beta)^{2}},
\end{align*}
where the last term simplifies to
\begin{align*}
 \frac{1}{2}\left[1-c^{ 2}(\lambda)\right]^{\frac{1}{2}} \sqrt{\|\tilde\beta\|^{2}\|x_{0}\|^{2} - (x_{0}^{\T}\tilde\beta)^{2}} =  
 x_{0}^{\T}\tilde\beta
\frac{\lambda(\|\tilde\beta\|^{ 2}-(x_{0}^{\T}\tilde\beta)^{2}/\|x_{0}\|^{ 2})}{
\|\tilde\beta\|^{2}\sqrt{(\|\tilde\beta\|^{ 2}+\lambda)^{ 2}-4\lambda(x_{0}^{\T}\tilde\beta)^{2}/\|x_{0}\|^{ 2}}}. 
\end{align*}
Hence
\begin{align*}
x_{0}^{\T}\hat \beta(\lambda)  
%
%
=& x_{0}^{\T}\tilde\beta\left[\frac{1}{2}+ \frac{1}{2}\frac{\|\tilde\beta\|^{2}(\|\tilde\beta\|^{ 2} + \lambda) -2\lambda(x_{0}^{\T}\tilde\beta)^{ 2}/\|x_{0}\|^{ 2}}{
\|\tilde\beta\|^{2}\sqrt{(\|\tilde\beta\|^{ 2}+\lambda)^{ 2}-4\lambda(x_{0}^{\T}\tilde\beta)^{2}/\|x_{0}\|^{ 2}}}\right] 
- 
\frac{x_{0}^{\T}\tilde\beta\lambda(\|\tilde\beta\|^{ 2}-(x_{0}^{\T}\tilde\beta)^{2}/\|x_{0}\|^{ 2})}{
\|\tilde\beta\|^{2}\sqrt{(\|\tilde\beta\|^{ 2}+\lambda)^{ 2}-4\lambda(x_{0}^{\T}\tilde\beta)^{2}/\|x_{0}\|^{ 2}}}, \\ 
=& x_{0}^{\T}\tilde\beta \left[\frac{1}{2}+\frac{1}{2}\frac{\|\tilde\beta\|^{2} - \lambda}{\sqrt{(\|\tilde\beta\|^{ 2}+\lambda)^{ 2}-4\lambda(x_{0}^{\T}\tilde\beta)^{2}/\|x_{0}\|^{ 2}}}\right].
\end{align*}

\subsection{Proof of Lemma~\ref{lem:derivative}}
Assuming a scaled orthogonal design, $X^{\T}X = nI_{p}$, introduces a scaling of the tuning parameters of the PAN-ridge prediction, denoted by $\hat\mu_{0}(\lambda_{1},\lambda_{2}) = x_{0}^{\T}\hat \beta(\lambda_{1},\lambda_{2})$
\begin{align*}
\hat\mu_{0}(\lambda_{1},\lambda_{2})  =  \frac{x_{0}^{\T}\tilde\beta}{1+\lambda_{1}/n}\left[\frac{1}{2}+\frac{1}{2}\frac{\|\tilde\beta\|^{2}-(1+\lambda_{1}/n)(\lambda_{2}/n)}{ \sqrt{(\|\tilde\beta\|^{2}+(1+\lambda_{1}/n)(\lambda_{2}/n))^{2} - 4(1+\lambda_{1}/n)(\lambda_{2}/n)(x_{0}^{\T}\tilde\beta)^{2}/\|x_{0}\|^{2}}}\right].
\end{align*}
The derivative of the mean squared error (MSE) of the prediction, denoted $\mu_{0}= x_{0}^{\T}\beta$, is bounded in a neighborhood of 0, such that
\begin{align*} 
\left.\frac{\partial\mse(x_{0},\beta, \lambda_{1},\lambda_{2})}{\partial\lambda_{2}}\right|_{\lambda_{2}=0} 
&= E\left[  \left. 2 (\hat\mu_{0}(\lambda_{1},0) - \mu_{0})  \frac{\partial\hat\mu_{0}(\lambda_{1},\lambda_{2})}{\partial\lambda_{2}}\right|_{\lambda_{2}=0} \mid X\right],
\end{align*}
where the derivative is 
\begin{align*}
   \frac{\partial\hat\mu_{0}(\lambda_{1},\lambda_{2})}{\partial\lambda_{2}}
&= -  \frac{x_{0}^{\T}\tilde\beta\left(\|x_{0}\|^{2}\|\tilde\beta\|^{2} - (x_{0}^{\T}\tilde\beta)^{2}\right) \left((1+\lambda_{1}/n)(\lambda_{2}/n) + \|\tilde\beta\|^{2}\right)}
{ n \|x_{0}\|^{2}\left[(\|\tilde\beta\|^{2}+(1+\lambda_{1}/n)(\lambda_{2}/n))^{2} - 4(1+\lambda_{1}/n)(\lambda_{2}/n)(x_{0}^{\T}\tilde\beta)^{2}/\|x_{0}\|^{2}\right]^{3/2}}. 
\end{align*}
The derivative of the MSE evaluated at $\lambda_2=0$ is given 
\begin{align*} 
\left.\frac{\partial\mse(x_{0},\beta, \lambda_{1},\lambda_{2})}{\partial\lambda_{2}}\right|_{\lambda_{2}=0} 
&= - \frac{2}{n} E\left[f(\tilde\beta)\right],
 \end{align*} 
 where 
\begin{equation}
f(\tilde\beta) = \left( x_{0}^{\T}\tilde\beta 
 - (1+\lambda_{1}/n)x_{0}^{\T}\beta\right) \frac{x_{0}^{\T}\tilde\beta (\|x_{0}\|^{2}\|\tilde\beta\|^{2}-(x_{0}^{\T}\tilde\beta)^{2})}{(1+\lambda_{1}/n)\|x_{0}\|^{2}\|\tilde\beta\|^{ 4}}. \label{function}
\end{equation}
Under the scaled orthogonal design, the variance of the OLS estimate is $ \var (\tilde\beta) = \sigma^{ 2}/n$, and the estimator converges in distribution as $ \sqrt{n}(\tilde\beta -\beta) \to \mathcal{N}(0,\sigma^{ 2})$, such that the expectation of the Taylor expansion of a function of the estimator is 
\begin{align*} E\left[f(\tilde\beta)\right] = f(\beta) +  \frac{1}{2} \frac{\sigma^{ 2}}{n} \tr(\mathbf{H}(f(\beta))) + O(1/n^{2}),
\end{align*} where $\mathbf{H}$ is the Hessian and $\tr(\mathbf{H}(f(\beta)))$ equals the Laplacian evaluated at $\beta$. As the Laplacian of \eqref{function} is
$$ \nabla^{ 2} f(\beta) = 
-\frac{2\left((\|x_{0}\|^{2}\|\beta\|^{2}-4(x_{0}^{\T}\beta)^{2})(\|x_{0}\|^{2}\|\beta\|^{2}-(x_{0}^{\T}\beta)^{2}) + (\lambda_{1}/n)(x_{0}^{\T}\beta)^{2}(3\|x_{0}\|^{2}\|\beta\|^{2}-4(x_{0}^{\T}\beta)^{2}) 
\right)}{(1+\lambda_{1}/n)\|x_{0}\|^{2}\|\beta\|^{6}}, $$
the expectation is given 
\begin{align*} 
&E\left[f(\tilde\beta)\right] =
( x_{0}^{\T}\beta - (1+\lambda_{1}/n)x_{0}^{\T}\beta) \frac{x_{0}^{\T}\beta (\|x_{0}\|^{2}\|\beta\|^{2}-(x_{0}^{\T}\beta)^{2})}{(1+\lambda_{1}/n)\|x_{0}\|^{2}\|\beta\|^{ 4}}  
\\ 
& + \frac{\sigma^{ 2}}{n} \frac{(\|x_{0}\|^{2}\|\beta\|^{2}-4(x_{0}^{\T}\beta)^{2})(\|x_{0}\|^{2}\|\beta\|^{2}-(x_{0}^{\T}\beta)^{2}) + (\lambda_{1}/n)(x_{0}^{\T}\beta)^{2}(3\|x_{0}\|^{2}\|\beta\|^{2}-4(x_{0}^{\T}\beta)^{2})}{(1+\lambda_{1}/n)\|x_{0}\|^{2}\|\beta\|^{ 6}} + O\left(\frac{1}{n^{2}}\right), \\
& =  -\frac{(\|x_{0}\|^{2}\|\beta\|^{2}-(x_{0}^{\T}\beta)^{2})}{n(1+ \lambda_{1}/n)\|x_{0}\|^{2}\|\beta\|^{ 4}} \bigg( \lambda_{1}(x_{0}^{\T}\beta)^{2} 
 -  \sigma^{ 2}\|x_{0}\|^{2} \Big(1-4 \frac{(x_{0}^{\T}\beta)^{2}}{\|x_{0}\|^{2}\|\beta\|^{2}}\Big) \bigg) + O\left(\frac{1}{n^{2}}\right).
\end{align*}
The expectation of the limit is hence given
\begin{align}
\hspace{-13pt}\left.\frac{\partial\mse(x_{0},\beta, \lambda_{1},\lambda_{2})}{\partial\lambda_{2}}\right|_{\lambda_{2}=0} & = C_1\bigg( \lambda_{1}(x_{0}^{\T}\beta)^{2} 
 -  \sigma^{ 2}\|x_{0}\|^{2} \Big(1-4 \frac{(x_{0}^{\T}\beta)^{2}}{\|x_{0}\|^{2}\|\beta\|^{2}}\Big)\bigg) + O\left(\frac{1}{n^{3}}\right), \label{eq:MSEderivative}
\end{align}
with $C_1 =  2(\|x_{0}\|^{2}\|\beta\|^{2}-(x_{0}^{\T}\beta)^{2})/(n(n+ \lambda_{1})\|x_{0}\|^{2}\|\beta\|^{ 4})$. 

\subsection{Proof of Theorems~\ref{thm:uniform} and~\ref{thm:uniform2}}
For $\lambda_{1}=0$, as the constant $C_1$ in \eqref{eq:MSEderivative} is always positive, the limit of the derivative will satisfy asymptotically as $\lambda_1 = 0$ 
$$ \lim_{n\to\infty}n^2\left.\frac{\partial\mse(x_{0},\beta, 0,\lambda_{2})}{\partial\lambda_{2}}\right|_{\lambda_{2}=0} \begin{cases} < 0, & \mbox{if } |x_{0}^{\T}\beta| < \frac{1}{2}\|x_{0}\||\beta\|, \\ > 0, & \mbox{if } |x_{0}^\T\beta| > \frac{1}{2}\|x_{0}\||\beta\|. \end{cases}$$
For $\lambda_{1}>0$, Equation \eqref{eq:MSEderivative} fulfills $\left.\frac{\partial\mse(x_{0},\beta, \lambda^{*}_{1},\lambda_{2})}{\partial\lambda_{2}}\right|_{\lambda_{2}=0} = 0$ for $$\lambda_{1}^{*} =  \frac{\sigma^{ 2}\|x_{0}\|^{2}}{(x_{0}^{\T}\beta)^{ 2}} 
\Big(1-4 \frac{(x_{0}^{\T}\beta)^{2}}{\|x_{0}\|^{2}\|\beta\|^{2}}\Big).$$ For $\lambda_{1} \neq \lambda_{1}^{*}$, the limit of the derivative therefore satisfies asymptotically
$$ \lim_{n\to\infty}n^2\left.\frac{\partial\mse(x_{0},\beta, \lambda_{1},\lambda_{2})}{\partial\lambda_{2}}\right|_{\lambda_{2}=0}  \begin{cases} < 0, & \mbox{if } 0 < \lambda_{1} < \lambda_{1}^{*} , \\ > 0, & \mbox{if } \lambda_{1} > \lambda_{1}^{*}. \end{cases}$$
\bibliographystyle{chicago}
\bibliography{pan_shortjournal}

\begin{thebibliography}{}

\bibitem[\protect\citeauthoryear{Cai, Fan, and Jiang}{Cai
  et~al.}{2013}]{cai2013distributions}
Cai, T., J.~Fan, and T.~Jiang (2013).
\newblock Distributions of angles in random packing on spheres.
\newblock {\em J. Mach. Learn. Res.\/}~{\em 14\/}(1), 1837--1864.

\bibitem[\protect\citeauthoryear{Cama and Harrison}{Cama and
  Harrison}{2018}]{cama2018fraud}
Cama, K.~J. and D.~T. Harrison (2018, June~26).
\newblock Fraud detection employing personalized fraud detection rules.
\newblock US Patent App. 10/007,914.

\bibitem[\protect\citeauthoryear{Carri{\'o}n, Cornblatt, Burton, Tso, Auther,
  Adelsheim, Calkins, Carter, Niendam, Sale, et~al.}{Carri{\'o}n
  et~al.}{2016}]{carrion2016personalized}
Carri{\'o}n, R.~E., B.~A. Cornblatt, C.~Z. Burton, I.~F. Tso, A.~M. Auther,
  S.~Adelsheim, R.~Calkins, C.~S. Carter, T.~Niendam, T.~G. Sale, et~al.
  (2016).
\newblock Personalized prediction of psychosis: external validation of the
  {NAPLS}-2 psychosis risk calculator with the {EDIPPP} project.
\newblock {\em Am. J. Psychiatry\/}~{\em 173\/}(10), 989--996.

\bibitem[\protect\citeauthoryear{Cheng, Alexander, MacLennan, Cummings,
  Montironi, Lopez-Beltran, Cramer, Davidson, and Zhang}{Cheng
  et~al.}{2012}]{cheng2012molecular}
Cheng, L., R.~E. Alexander, G.~T. MacLennan, O.~W. Cummings, R.~Montironi,
  A.~Lopez-Beltran, H.~M. Cramer, D.~D. Davidson, and S.~Zhang (2012).
\newblock Molecular pathology of lung cancer: key to personalized medicine.
\newblock {\em Mod. Pathol.\/}~{\em 25\/}(3), 347.

\bibitem[\protect\citeauthoryear{Cho}{Cho}{2009}]{cho2009inner}
Cho, E. (2009).
\newblock Inner product of random vectors.
\newblock {\em Int. J. Pure Appl. Math.\/}~{\em 56\/}(2), 217--221.

\bibitem[\protect\citeauthoryear{Claeskens and Hjort}{Claeskens and
  Hjort}{2003}]{claeskens2003focused}
Claeskens, G. and N.~L. Hjort (2003).
\newblock The focused information criterion.
\newblock {\em J. Am. Statist. Assoc.\/}~{\em 98\/}(464), 900--916.

\bibitem[\protect\citeauthoryear{Claeskens and Hjort}{Claeskens and
  Hjort}{2008}]{claeskens2008model}
Claeskens, G. and N.~L. Hjort (2008).
\newblock {\em Model selection and model averaging}.
\newblock Cambridge University Press.

\bibitem[\protect\citeauthoryear{Cohl}{Cohl}{2011}]{cohl2011laplace}
Cohl, H.~S. (2011).
\newblock Opposite antipodal fundamental solution of {L}aplace's equation in
  {H}yperspherical geometry.
\newblock {\em Symmetry Integr. Geom.\/}~{\em 7}, 108--122.

\bibitem[\protect\citeauthoryear{Efron and Tibshirani}{Efron and
  Tibshirani}{1994}]{efron1994introduction}
Efron, B. and R.~J. Tibshirani (1994).
\newblock {\em An introduction to the bootstrap}.
\newblock CRC press.

\bibitem[\protect\citeauthoryear{Fan, Han, and Liufa}{Fan
  et~al.}{2014}]{fan2014challenges}
Fan, J., F.~Han, and H.~Liufa (2014).
\newblock Challenges of {B}ig data analysis.
\newblock {\em Natl. Sci. Rev.\/}~{\em 1\/}(2), 293--314.

\bibitem[\protect\citeauthoryear{Gatto and Jammalamadaka}{Gatto and
  Jammalamadaka}{2007}]{gatto2007generalized}
Gatto, R. and S.~R. Jammalamadaka (2007).
\newblock The generalized von mises distribution.
\newblock {\em Stat. Methodol.\/}~{\em 4\/}(3), 341--353.

\bibitem[\protect\citeauthoryear{Hamburg and Collins}{Hamburg and
  Collins}{2010}]{hamburg2010path}
Hamburg, M.~A. and F.~S. Collins (2010).
\newblock The path to personalized medicine.
\newblock {\em New Engl. J. Med.\/}~{\em 363\/}(4), 301--304.

\bibitem[\protect\citeauthoryear{Hellton and Hjort}{Hellton and
  Hjort}{2018}]{hellton2018fridge}
Hellton, K.~H. and N.~L. Hjort (2018).
\newblock Fridge: Focused fine-tuning of ridge regression for personalized
  predictions.
\newblock {\em Stat. Med.\/}~{\em 37\/}(8), 1290--1303.

\bibitem[\protect\citeauthoryear{Hoerl and Kennard}{Hoerl and
  Kennard}{1970}]{hoerl1970ridge}
Hoerl, A.~E. and R.~W. Kennard (1970).
\newblock Ridge regression: Biased estimation for nonorthogonal problems.
\newblock {\em Technometrics\/}~{\em 12\/}(1), 55--67.

\bibitem[\protect\citeauthoryear{Huang, D{\"u}ren, Hellton, and Lederer}{Huang
  et~al.}{2019}]{huang2019tuning}
Huang, S.-T., Y.~D{\"u}ren, K.~H. Hellton, and J.~Lederer (2019).
\newblock Tuning parameter calibration for prediction in personalized medicine.
\newblock {\em arXiv preprint arXiv:1909.10635\/}.

\bibitem[\protect\citeauthoryear{Kosorok and Laber}{Kosorok and
  Laber}{2019}]{kosorok2019precision}
Kosorok, M.~R. and E.~B. Laber (2019).
\newblock Precision medicine.
\newblock {\em Annu. Rev. Stat. Appl.\/}~{\em 6}, 263--286.

\bibitem[\protect\citeauthoryear{Liu and Meng}{Liu and
  Meng}{2016}]{liu2016there}
Liu, K. and X.-L. Meng (2016).
\newblock There is individualized treatment. why not individualized inference?
\newblock {\em Annu. Rev. Stat. Appl.\/}~{\em 3}, 79--111.

\bibitem[\protect\citeauthoryear{Liu, Zhang, Li, Yu, Dai, Zhao, and Song}{Liu
  et~al.}{2017}]{liu2017deep}
Liu, W., Y.-M. Zhang, X.~Li, Z.~Yu, B.~Dai, T.~Zhao, and L.~Song (2017).
\newblock Deep hyperspherical learning.
\newblock In {\em Advances in Neural Information Processing Systems}, pp.\
  3950--3960.

\bibitem[\protect\citeauthoryear{Mardia}{Mardia}{1972}]{mardia1972statistics}
Mardia, K.~V. (1972).
\newblock {\em Statistics of directional data}.
\newblock New York: Academic press.

\bibitem[\protect\citeauthoryear{{\"O}hrn and Linderberg}{{\"O}hrn and
  Linderberg}{1983}]{ohrn1983hyperspherical}
{\"O}hrn, Y. and J.~Linderberg (1983).
\newblock Hyperspherical coordinates in four particle systems.
\newblock {\em Mol. Phys.\/}~{\em 49\/}(1), 53--64.

\bibitem[\protect\citeauthoryear{Pourahmadi and Wang}{Pourahmadi and
  Wang}{2015}]{pourahmadi2015distribution}
Pourahmadi, M. and X.~Wang (2015).
\newblock Distribution of random correlation matrices: Hyperspherical
  parameterization of the cholesky factor.
\newblock {\em Stat. Probabil. Lett.\/}~{\em 106}, 5--12.

\bibitem[\protect\citeauthoryear{Rafailidis, Axenopoulos, Etzold, Manolopoulou,
  and Daras}{Rafailidis et~al.}{2014}]{rafailidis2014content}
Rafailidis, D., A.~Axenopoulos, J.~Etzold, S.~Manolopoulou, and P.~Daras
  (2014).
\newblock Content-based tag propagation and tensor factorization for
  personalized item recommendation based on social tagging.
\newblock {\em ACM Trans. Interact. Intell. Syst.\/}~{\em 3\/}(4), 26.

\bibitem[\protect\citeauthoryear{Reber, Canning, and Harackiewicz}{Reber
  et~al.}{2018}]{reber2018personalized}
Reber, R., E.~A. Canning, and J.~M. Harackiewicz (2018).
\newblock Personalized education to increase interest.
\newblock {\em Current directions in psychological science\/}~{\em 27\/}(6),
  449--454.

\bibitem[\protect\citeauthoryear{Scealy and Welsh}{Scealy and
  Welsh}{2011}]{scealy2011regression}
Scealy, J.~L. and A.~H. Welsh (2011).
\newblock Regression for compositional data by using distributions defined on
  the hypersphere.
\newblock {\em J. R. Statist. Soc. B\/}~{\em 73\/}(3), 351--375.

\bibitem[\protect\citeauthoryear{Stamey, Kabalin, McNeal, Johnstone, Freiha,
  Redwine, and Yang}{Stamey et~al.}{1989}]{stamey1989prostate}
Stamey, T.~A., J.~N. Kabalin, J.~E. McNeal, I.~M. Johnstone, F.~Freiha, E.~A.
  Redwine, and N.~Yang (1989).
\newblock Prostate specific antigen in the diagnosis and treatment of
  adenocarcinoma of the prostate. {II}. radical prostatectomy treated patients.
\newblock {\em J. Urol.\/}~{\em 141\/}(5), 1076--1083.

\bibitem[\protect\citeauthoryear{Tang, Liao, and Sun}{Tang
  et~al.}{2013}]{tang2013prediction}
Tang, H., S.~S. Liao, and S.~X. Sun (2013).
\newblock A prediction framework based on contextual data to support mobile
  personalized marketing.
\newblock {\em Decis. Support Syst.\/}~{\em 56}, 234--246.

\bibitem[\protect\citeauthoryear{Tian and Zhao}{Tian and
  Zhao}{2015}]{tian2015statistical}
Tian, L. and X.~Zhao (2015).
\newblock Statistical methods for personalized medicine.
\newblock In Y.~Lu, J.-q. Fang, L.~Tian, and J.~Hua (Eds.), {\em Advanced
  medical statistics\/} (2nd ed.)., pp.\  79--102. World Scientific.

\bibitem[\protect\citeauthoryear{Tibshirani}{Tibshirani}{1996}]{tibshirani1996regression}
Tibshirani, R. (1996).
\newblock Regression shrinkage and selection via the lasso.
\newblock {\em J. R. Statist. Soc. B\/}~{\em 58\/}(1), 267--288.

\bibitem[\protect\citeauthoryear{Van~der Laan and Rose}{Van~der Laan and
  Rose}{2011}]{van2011targeted}
Van~der Laan, M.~J. and S.~Rose (2011).
\newblock {\em Targeted learning: causal inference for observational and
  experimental data}.
\newblock Berlin: Springer.

\bibitem[\protect\citeauthoryear{Zeevi, Korem, Zmora, Israeli, Rothschild,
  Weinberger, Ben-Yacov, Lador, Avnit-Sagi, Lotan-Pompan, et~al.}{Zeevi
  et~al.}{2015}]{zeevi2015personalized}
Zeevi, D., T.~Korem, N.~Zmora, D.~Israeli, D.~Rothschild, A.~Weinberger,
  O.~Ben-Yacov, D.~Lador, T.~Avnit-Sagi, M.~Lotan-Pompan, et~al. (2015).
\newblock Personalized nutrition by prediction of glycemic responses.
\newblock {\em Cell\/}~{\em 163\/}(5), 1079--1094.

\bibitem[\protect\citeauthoryear{Zhang and Nebert}{Zhang and
  Nebert}{2017}]{zhang2017personalized}
Zhang, G. and D.~W. Nebert (2017).
\newblock Personalized medicine: Genetic risk prediction of drug response.
\newblock {\em Pharmacol. Therapeut.\/}~{\em 175}, 75--90.

\bibitem[\protect\citeauthoryear{Zou and Hastie}{Zou and
  Hastie}{2005}]{zou2005regularization}
Zou, H. and T.~Hastie (2005).
\newblock Regularization and variable selection via the elastic net.
\newblock {\em J. R. Statist. Soc. B\/}~{\em 67\/}(2), 301--320.

\end{thebibliography}

\end{document}